\documentstyle[art10]{article}
\begin{document}

\title {Continuum field theory of string-like objects.
Dislocations and superconducting vortices}
\author {Dominik Rogula\thanks{Laboratory for Physics of
Structured Matter,
Institute for Fundamental Technological Research,
ul. \a'{S}wi\c{e}tokrzyska 21, PL-00-049 Warsaw.
E--mail: drogula@ippt.gov.pl}}
\date{}
\maketitle

\newcommand{\equn}[1]{\begin{equation} {#1} \end{equation}}
\newcommand{\equln}[2]{\begin{equation} {#1} \label{#2} \end{equation}}

\hfill To the memory of Ekkehart Kr\"{o}ner\\

{\footnotesize {\bf Summary:}
Dense distributions of string-like objects in material media are
considered in terms of continuum field theory. The strings are
assumed to carry a quantized abelian topological charge, such as
the Burgers vector of dislocations in solids, or magnetic flux of
supercurrent vortices in type-II superconductors.
Within this common framework
the facts known from dislocation theory can be extended, in
appropriately modified forms, to other physical contexts.
In particular, the concept of incompatible distortions is
transplanted into the theory of type-II superconductivity.
The compatibility law for type-II
superconductors is derived in terms of differential forms.
As a result, one obtains an inhomogeneous generalization of
the classic Londons' equation.

\normalsize
\section{Introduction}

The heuristics of the present paper can be conveniently developed
by starting from the striking analogy between dislocations in solids
and magnetic vortices in superconductors.  While the former
determine the behaviour of materials in plastic deformation
processes, the latter are responsible for magnetic and transport
phenomena in type-II superconductors. The physical contexts
in the situations are indeed very different.
Nevertheless, the analogy
is far from being superficial. It suffices to note common fundamental
features such as the distinguished linear structure, quantized
dynamical quantities concentrated on and conserved along the lines,
as well as large amount of irreversibility in their dynamical behaviour.

    In most situations the number of vortices in
a macroscopic superconducting specimen is very large, what enables
an effective description of vortex networks with the aid of continuum
field theories [1-3, 5--6, 8--11, 18].

   In the present paper we shall concentrate our attention on
the compatibility laws which govern the behaviour of the continuum
fields describing the macroscopic quantities. We adopt the
following terminological convention. Whenever more precise language
concerning type II superconductivity is necessary
we shall make use of the term {\em fluxon} in place of
informal terms like {\em flux line} or {\em magnetic vortex}.
It, however, should be contrasted with the term {\em fluxoid},
for which we adopt London's definition \cite{London I}.

   Thoughout the paper, a theoretical description of a string system
will be termed {\em macroscopic} if it deals with collective
quantities without paying attention to individual lines. The
description which refers to individual strings, but not to
individual atoms, will be called {\em mesoscopic};
the term {\em microscopic} will be reserved
for the atomistic desription.


\section{Analogy with dislocation networks}\label{Analogy}
In general terms, the relation between magnetic flux lines
and macroscopic magnetic properties of superconductors is
similar to the relation between dislocation lines and
the macroscopic mechanical properties of solids.

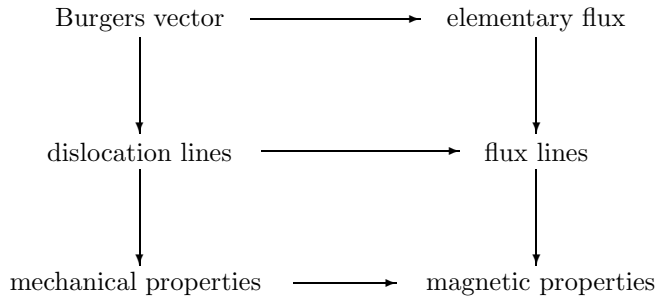
\vspace{10pt}
\begin{picture}(320,114)
\put(80,107){\makebox(0,0){Burgers vector}}
\put(230,107){\makebox(0,0){elementary flux}}
\put(80,57){\makebox(0,0){dislocation lines}}
\put(230,57){\makebox(0,0){flux lines}}
\put(80,7){\makebox(0,0){mechanical properties }}
\put(230,7){\makebox(0,0){ magnetic properties}}
\put(154,110){\makebox(0,0)[b]{$\scriptstyle $}}\put(122,107){\vector(1,0){64}}
\put(230,82){\makebox(0,0)[l]{$\scriptstyle $}}\put(230,100){\vector(0,-1){36}}
\put(80,82){\makebox(0,0)[l]{$\scriptstyle $}}\put(80,100){\vector(0,-1){36}}
\put(163,60){\makebox(0,0)[b]{$\scriptstyle $}}\put(126,57){\vector(1,0){75}}
\put(230,32){\makebox(0,0)[l]{$\scriptstyle $}}\put(230,50){\vector(0,-1){36}}
\put(80,32){\makebox(0,0)[l]{$\scriptstyle $}}\put(80,50){\vector(0,-1){36}}
\put(157,10){\makebox(0,0)[b]{$\scriptstyle $}}\put(138,7){\vector(1,0){39}}
\end{picture}
\vspace{10pt}

\noindent
Later on we shall
find limitations of this analogy. For the present purposes it
turns out to be useful, and we shall try to get some profits from it.

   The continuum theory of distributed dislocations
has been developed by Kr\"{o}ner in his classic work
\cite{Kroner kontinuumstheorie}.
The theory has been generalized to moving dislocations
by Kosevitch \cite{Kosevitch 79}.
Modern geometric formulations of continuum theory of
dislocations have been given by Kr\"{o}ner \cite{Kroner 81, Kroner 89},
Mistura \cite{Mistura}, and Trz\c{e}sowski \cite{Trzesowski 87}.
The dynamics of networks of string-like objects has been considered
recently by Rogula \& Sztyren \cite{Rogula+Sztyren 2000,
Rogula+Sztyren 2001}.

   A single dislocation line L is characterized by its Burgers
vector \(b^i_L\), which is topologically quantized and conserved
along the line. On the other hand, a single magnetic flux line
is characterized by associated magnetic flux \(\Phi_L\), which
also is topologically quantized and conserved along the line.
An immediately visible difference between the two cases is the
vectorial nature of \(b^i_L\) and the pseudoscalar nature of \(\Phi_L\).
This difference, however, athough important in many respects,
does not break the analogy.

\normalsize
   The presence of dislocation lines in a material continuum
results in a distortion field which is well defined in the region
of "good material" -- outside the dislocation cores. The good
material is, however, multi-connected and the dislocations give
rise to the distortion field which cannot be represented
with the aid
of a globally defined single--valued displacement field.
One of the possibilities here is to make use of
multi--valued displacement fields.
Another possibility consists in the following.
Instead of a multi--valued displacement field
\(u^i\), one considers a single--valued differential form \(du^i\)
defined by the relation\
   \equln {\oint_C du^i=b^i[C],} {dui}
\normalsize \\
where the right--hand side denotes the total Burgers vector of
the dislocation lines encircled by the contour \(C\).
The entire contour \(C\) is confined to the good material
and is otherwise arbitrary.

\section{Macroscopic averages in string systems}

To describe effectively the material media which contain a
very large amount of string-like objects
one has the need for a fully
macroscopic theory which operates with quantities describing large
collections of such objects without paying much attention
to an individual string.

\subsection{String density and string current}\label{Strid}

   The placement and movement of an individual line segment 
at the
mesoscopic level can be conveniently described with the aid of the
following quantities,
valid generally for networks of moving lines \(L=L(t)\)
\equn{\alpha^j(x\mid L)=\int_L\delta({\bf x}-{\bf x^{'}})dx^{'j}}
\normalsize
and
\equln{\iota_k(x,t\mid L(t))=\int_{L(t)}\delta({\bf x}-{\bf x^{'}})
      \epsilon_{kjl}v^l({\bf x^{'}})dx^{'j},}{jota}
\normalsize
where \(v^l({\bf x^{'}})\) denotes the velocity of the line element
at \({\bf x^{'}}\).  Only the components of velocity perpendicular to
the line element contribute to the string current (\ref{jota}).

   Note that 
the above defined quantities, which may be interpreted as the
{\em oriented string density} and {\em oriented string current},
depend solely on the orientation and geometry of the
lines, and are independent of any physical characteristics of the
strings.
   For the string density and string current we obtain the following
balance equations
\equn{\frac{\partial{\alpha^i}}{\partial{t}}+\epsilon^{ijk}\iota_{k,j}=0,
        \,\,\,\,\,\,\alpha^i_{\ ,i}=0.}
\normalsize
It is expedient to notice that, in the case of point-like particles,
the analogue of the first of the above equations expresses the
conservation of the number of particles.

   For networks of dislocations in crystals and magnetic flux lines
in superconductors, the related quatities are defined in the
following subsections.

\subsection{Dislocation networks}
 Notation: \(b^i_L\) represents the Burgers
vector associated with the line segment \(L\).
\equn{\alpha^{ij}(x\mid L)=b_L^i\alpha^j(x\mid L),}
\equn{\alpha^{ij}_{meso}(x)=\sum_L\alpha^{ij}(x\mid L ),}
\equln{\alpha^{ij}(x)=\langle\alpha^{ij}_{meso}(x)\rangle_{av},}
   {alfaij}
\equn{J_{\ k}^{i\ \ meso}(x)=\sum_{L(t)}b^{i}_L\iota_k(x,t\mid L(t)),}
\equn{J^i_{\ k}(x,t)=\langle J_{\ k}^{i\ \ meso}(x,t)\rangle_{av}.}
\normalsize
The dislocation density \(\alpha^{ij}\) and the dislocation current
\(J^i_{\ k}\) satisfy the equations
\equln{\frac{\partial}{\partial t}\alpha^{ij}+
   \epsilon^{jlk}\frac{\partial}{\partial x^l}J^i_{\ k}=0,
   \ \ \ \ \alpha^{ij}_{\ ,j}=0}
   {Kos}
\normalsize
which express the Burgers vector conservation laws.

\subsection{Magnetic flux networks}
Notation: \(\Phi_L\) and \(\Phi_0\) represent
the quantized flux associated with the line segment \(L\) and the
elementary flux quantum, respectively. Typically \(\Phi_L=\pm\Phi_0\).
\equn{\Theta^{j}(x\mid L)=\Phi_L\alpha^j(x\mid L).}
\equn{\Theta^{j}_{meso}(x)=\sum_L\Theta^{j}(x\mid L ),}
\equln{\Theta^{j}(x)=\langle \Theta^{j}_{meso}(x)\rangle_{av},}
    {Teta}
\equn{K_{i}^{meso}(x)=\sum_{L(t)}\Phi_L\iota{_i}(x,t\mid L(t)),}
\equn{K_{i}(x,t)=\langle K_{i}^{meso}(x,t)\rangle_{av}.}
\normalsize
The density \(\Theta_i\) and the current \(K_i\) satisfy 
the conservation equations
\equln{\frac{\partial}{\partial t}\Theta^{i}+
   \epsilon^{ilj}\frac{\partial}{\partial x^l}K_j=0,
   \ \ \ \ \Theta^{i}_{\,,i}=0}{eqI}
\normalsize
which are the analogues of eqns. (\ref {Kos}).
The quantity \(\Theta^i\) represents the contribution of the
magnetic flux lines (fluxons) to the macroscopic
induction field \(B^i\); therefore it will be called the
(oriented) {\em  fluxon density}.
The quantity \(K_j\) represents the macroscopic
{\em fluxon current} density
resulting from the motion of fluxons.
On the other hand, the flux current produces
some electric field, coupled to the moving magnetic flux
through Maxwell equations. In consequence, we arrive at
the following interpretation
\equn{B^{str}=\Theta,\ \ \ E^{str}=\frac{1}{c}K,}
\normalsize
where \(B^{str}\) and \(E^{str}\) stand for the contribution of
strings to the electromagnetic fields.

    Several approaches to the continuum description of
superconductors containing dense distributions of vortex lines
are known from the literature;
see e.g. London \& London \cite{London+London 35a, London I},
Laue \cite{Laue book}, Zhou \cite{Zhou book},
Abrikosov \cite {Abrikosov 57, Abrikosov book},
Anthony \& Seeger \cite {Anthony+Seeger 73},
Chapman {\em et al.}
\cite{Chapman+ 92, Chapman 2000},
Rogula \cite{Rogula 99}. Some of them can be derived with the
aid of particular assumptions concerning the macroscopic
averages. Note that, in general,
due to statistical correlations between
mesoscopic string configurations and their motions%
, the inequality
\equln{\langle \alpha^j v^l \rangle \neq
        \langle \alpha^j \rangle \langle v^l \rangle}{Neq}
\normalsize
%
holds. There are, however, important special cases, such as
coherent fluxon flow, or else uncorrelated
randomness of \(\alpha\) and \(v\), when both sides of the formula
(\ref{Neq}) are identical to a good degree of approximation,
and the corresponding assumption of the vortex-density model
\cite{Chapman 2000} is justified.
The fluxon current \(K\) and the fluxon
density \(\Theta\) are then related through
the macroscopic average of the fluxon's velocity by the equation
\equn{K+v\wedge\Theta=0,}
\normalsize
valid in this special case.

\section{The compatibility equations}
\subsection{The distortion field and dislocation density}

   The presence of continuously distributed dislocation lines
in a material continuum causes an incompatible macroscopic 
distortion of the medium. The situation
may be briefly sketched as follows. While the multi-valued mesoscopic
displacement fields, mentioned in Section \ref{Analogy}, are
rather hard to be averaged, and would lead to ill-defined
macroscopic diplacement fields, the single-valued differential
forms allow for unequivocal averages. Therefore, 
to describe incompatible distortion fields at the macroscopic level,
instead of ill--defined displacement field
\(u^i\) one considers the well--defined differential
form \(du^i\) undestood as the macroscopic average
of the mesoscopic differential form (\ref{dui}).
The corresponding continnum field equations, expressed in terms
of the material distortion \(\beta^i_j\) and the dislocation
density \(\alpha^{il}\), take the form
   \equln {\epsilon^{lkj}\,\beta^i_{j,k}=
   \alpha^{il},\ \ \ \alpha^{il}_{\ ,l} = 0,}{alfail}
\normalsize
and can be obtained from the expression
   \equln{du^i(x,t)=\beta^i_j(x,t)dx^j}{duixt}
\normalsize
by taking into account the equations (\ref{dui}) and (\ref{alfaij}).
Keeping that in mind, we will search for an analogue of the
above relations in the framework of superconductivity.

\subsection{The fluxoid density and superconducting vortices}
   Let us consider the expression
\equln{j_k=\frac{e^{*}\hbar}{2im^{*}}
           (\bar{\psi}\psi_{,k}-\bar{\psi}_{,k}\psi)
             -\frac{e^{*2}}{m^{*}c}\mid\psi\mid^2A_k.}{GL2}
\normalsize
for the supercurrent density (for simplicity we start
from an isotropic superconductor) with the following notation:
\(\psi=\psi(x,t)\) represents the Ginzburg-Landau order parameter,
\(\bar{\psi}\) its complex conjugate, \(A_k=A_k(x,t)\) the
electomagnetic vector potential, \(e^*\) and \(m^*\) denote
the electric charge
and the effective mass of a Cooper pair, respectively.
We substitute
\equn{\psi=\ \mid\psi\mid z,}
where \(\mid z\mid\ =1\) so that \(z=z(x,t)\) equals the
phase factor of the order parameter field. 
The eqn. (\ref{GL2}) may then be rewritten as
\equln{j_k=\frac{e^*}{m^*}\mid\psi\mid^2(\frac{\hbar}
{2i}(\bar{z}z,_k-\bar{z},_kz)-\frac{e*}{c}A_k).} {jotk}
\normalsize
Now, let us introduce the following differential form:
\equn{d\phi \stackrel{\rm def}{=}-i\bar{z}dz=i\,d\bar{z}z,}
\normalsize
which represents the phase differential.
With the aid of this form we can write
\equn{j_kdx^k=-\frac{e^*}{m^*}\mid\psi\mid^2(\hbar d\phi
+\frac{e^*}{c}A_kdx^k)}
\normalsize
or, taking into account that in the superconducting region
outside the vortices \(\psi\neq 0\),
\equn{-\frac{\hbar c}{e^*}d\phi=(A_k +
\frac{m^* c}{e^{* 2}}
\frac{j_k}{\mid\psi\mid^2})dx^k.}
\normalsize
The right--hand side of the above equation represents the
fluxoid density. After integration over an appropriately
smooth surface \(S\), we obtain
\equn{-\frac{\hbar c}{e^*}\oint_Cd\phi=\Phi[C],}
\normalsize
where \(\Phi [C]\) equals the total fluxoid encircled by
the contour \(C=\partial S\),
\equn{\Phi [C]=\int_S{\Theta^i}ds_i.}
\normalsize
%


   In consequence, passing to the differential
relation, we obtain the analogue of eqn. (\ref {alfail})
\equn{\epsilon^{ilk}(A_k+\frac{m^* c}{e^{* 2}}
\frac{j_k}{\mid\psi\mid^2}),_l=\Theta^i,\ \ \ \Theta^i_{\,,i}=0}
\normalsize
or equivalently
\equln{B^i + \,\epsilon^{ilk}(\frac{m^* c}
{e^{* 2}n_s} j_k),_l=\Theta^i,}{BThetaIso}
\normalsize
with \(n_s=\mid\psi\mid^2\).
The left--hand side of this equation corresponds to the
gauge invariant Londons' equation
\equln{B^i  +\,\epsilon^{ilk}\,(\frac{mc}{e^2n_s}\,j_k)_{,l}=0,
}{LondonsB}
\normalsize
with the charge
\(e\) and mass \(m\) replaced by the Cooper pair charge
\(e^*\) and effective mass \(m^*\), respectively,
and with \(\mid\psi\mid^2\) interpreted as \(n_s\).
Hence one can see that the incompatibility of the
superconducting order parameter due to continuously
distributed flux lines modifies Londons' relation
(\ref{LondonsB}) between
the magnetic induction field and the supercurrent density.

   Due to the electromagnetic character of the quantities involved,
an electric counterpart of eqn. (\ref{BThetaIso}) has also to be valid.
In fact, taking into account the flux flow balance (\ref{eqI})
and making use of the Maxwell equation
\equn{\epsilon^{ijk}E_{k,j}+\frac{1}{c}\frac{\partial B^i}
   {\partial t}=0,}
\normalsize
we obtain
\equn{\epsilon^{ilm}(\frac{\partial}{\partial t}
(\frac{m^* }{e^{* 2}n_s} j_m)+
 \frac{1}{c}K_m - E_m)_{,l}=0.}
\normalsize

The above equation can be conveniently integrated to the form
\equln{E_i = -\chi_{,i}+\frac{\partial}{\partial t}
(\frac{m^*}{e^{* 2}n_s} j_i)+
 \frac{1}{c}K_i.}{EchiIso}
\normalsize
Note, however, that the integrated relation (\ref{EchiIso}) becomes
algebraically independent of the equations derived up to this point.
This is due to the
new gradient term \(\chi_{,i}\) which depends on
specific features of the configuration under consideration.

    The equations given so far in the present section
are valid literally for isotropic
superconductors. The generalization to anisotropic superconductors
can be, however, performed in a straightforward manner by applying
the routine procedure based on substitution of the effective mass
tensor \(m^*_{ij}\) in place of the scalar mass \(m^*\).
The tensor \(m^*_{ij}\) is real and symmetric by definition, and
positive-definite by assumption; in particular, it has a well
defined inverse which can be substituted to eqn. (\ref{GL2}).
As a result, the equations (\ref{BThetaIso}) take the form
\equln{B^i + \,\frac{4\pi}{c}\epsilon^{ilj}(\lambda^2_{jk} j^k),_l=
       \Theta^i,}{BTheta}
\normalsize
and
\equln{E_i = -\chi_{,i}+\frac{4\pi}{c}\frac{\partial}{\partial t}
(\lambda^2_{ik} j^k) + \frac{1}{c}K_i.}{EchiAni}
\normalsize
The tensor
\equn{\lambda^2_{jk}=\frac{c^2}{4\pi}\frac{m^*_{jk}}
{e^{* 2}n_s}}
\normalsize
generalizes the (squared) Londons' penetration depth \(\lambda_L^2\).
%

\section{Conclusion}

    The heuristic formulation of the analogy between dislocations
in solids and flux lines in superconductors can now be stated in
a more precise way: the diagram given in the Introduction should
be complemented by the following scheme:

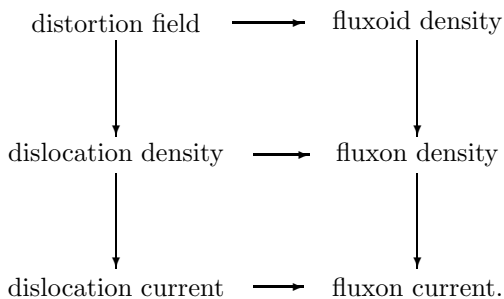
\vspace{10pt}
\begin{picture}(248,114)
\put(124,107){\makebox(0,0){distortion field}}
\put(238,107){\makebox(0,0){fluxoid density}}
\put(124,57){\makebox(0,0){dislocation density}}
\put(238,57){\makebox(0,0){fluxon density}}
\put(124,7){\makebox(0,0){dislocation current}}
\put(238,7){\makebox(0,0){fluxon current.}}
\put(181,110){\makebox(0,0)[b]{$\scriptstyle $}}\put(168,107){\vector(1,0){27}}
\put(238,82){\makebox(0,0)[l]{$\scriptstyle $}}\put(238,100){\vector(0,-1){36}}
\put(124,82){\makebox(0,0)[l]{$\scriptstyle $}}\put(124,100){\vector(0,-1){36}}
\put(186,60){\makebox(0,0)[b]{$\scriptstyle $}}\put(176,57){\vector(1,0){21}}
\put(238,32){\makebox(0,0)[l]{$\scriptstyle $}}\put(238,50){\vector(0,-1){36}}
\put(124,32){\makebox(0,0)[l]{$\scriptstyle $}}\put(124,50){\vector(0,-1){36}}
\put(185,10){\makebox(0,0)[b]{$\scriptstyle $}}\put(176,7){\vector(1,0){19}}
\end{picture}

\vspace{10pt}

   Due to the presence of extra quantities -- the fluxon density
and the fluxon current -- the above given macroscopic equations,
even when augmented with material constitutive
relations, are incomplete. They can be completed by a variety
of mathematical models stated in terms of the fields \(\Theta^i\)
and \(K_j\) and taking into account the string kinetics.
In this way one can, for instance, reproduce the
behaviour of superconductors described by critical state models,
such as Bean's \cite{Bean 62} or Kim \& Stephen's
\cite{Kim+Stephen 69} ones.
On the other hand,
by making use of the above stated analogy, one can adapt
a selection of models conceived in the theory of dislocations
in order to describe the irreversible behaviour of dense
string systems.

\section{Acknowledgements}
This work was supported by the Science Research
Committee (Poland) under grants No. 7 TO7A 010 16 and 5 TO7A 040 22.\\


\thebibliography{12}
\bibliography{}

\bibitem{Abrikosov 57}A.\,A. Abrikosov,
On magnetic properties of type II superconductors, {\em J. Exp. Th. Phys. (Russian)} {\bf 32}(1957)1442-1450 

\bibitem{Abrikosov book}A.\,A. Abrikosov,
{\em Fundamentals of the Theory of Metals,} North--Holland 1988  

\bibitem{Anthony+Seeger 73}K.\,H. Anthony and A. Seeger,
Eine nichtlineare Kontinuumstheorie des Flussliniengitters in Supraleitern, {\em Phil. Mag.} {\bf 28}(1973)1125-1148 

\bibitem{Bean 62}C.\,P. Bean,
Magnetization of hard superconductors, {\em Phys. Rev. Lett.} {\bf 8}(1962)250-253; {\em Rev. Mod. Phys.} {\bf 36}(1964)31-39 

\bibitem{Chapman+ 92}S.\,J. Chapman and S.\,D. Howson and J.R. Ockendon,
Macroscopic models of superconductivity, {\em SIAM Rev.} {\bf 34}(1992)529-560 

\bibitem{Chapman 2000}S.\,J. Chapman,
A hierarchy of models for type II superconductors, {\em SIAM Rev.} {\bf 42}(2000)555-598 

\bibitem{Kim+Stephen 69}Y.\,B. Kim and M.J. Stephen,
Flux flow and irreversible effects, {\em In: Superconductivity,} ed. H.Parks, New York 1969  

\bibitem{Laue book}M. von Laue,
{\em Theorie der Supraleitung,} Springer Verlag, Berlin-G\"{o}ttingen-Heidelberg 1949  

\bibitem{London+London 35a}F. London and H. London,
The electromagnetic equations of the supraconductor, {\em Proc. Roy. Soc. (London)} {\bf A149}(1935)71-88 

\bibitem{London I}F. London,
{\em Superfluids. I. Macroscopic theory of superconductivity,} Dover Publications, New York 1961  

\bibitem{Zhou book}
Shu-Ang Zhou, {\em Electrodynamic theory of superconductors}, Peter Peregrinus Ltd., London 1991  

\bibitem{Kosevitch 79}A.\,M. Kosevitch,
Crystal dislocations and the theory of elasticity, {\em Dislocations in Solids,} Vol. 1, ed. F.R.Nabarro, North--Holland 1979  

\bibitem{Kroner kontinuumstheorie}E. Kr\"{o}ner,
Kontinuumstheorie der Versetzungen und Eigenspannungen, {\em Ergebnise der Angewandte Mathematik} {\bf 5}, eds. L.Collatz, F.L\"{o}sch, Springer Verlag, Heidelberg 1958  

\bibitem{Kroner 81}E. Kr\"{o}ner,
Continuum theory of defects, {\em In: Physics of Defects,} Les Houches 1980, edited by R.Balian et al., North--Holland 1981  

\bibitem{Kroner 89}E. Kr\"{o}ner,
The differential geometry of elementary point and line defects in Bravais crystals, lecture at the Summer School on Topology, Geometry and Gauging, Jab{\l}onna 1989; {\em Int. J. Theor. Phys.} {\bf 29}(1990)1219-1237 


\bibitem{Mistura}L. Mistura,
Cartan connections and defects in Bravais lattices, lecture at the Summer School on Topology, Geometry and Gauging, Jab{\l}onna 1989; {\em Int. J. Theor. Phys.} {\bf 29}(1990)1207-1218 

\bibitem{Trzesowski 87}A. Trz\c{e}sowski,
Geometry of crystal structure with defects, {\em Int. J. Theor. Phys.} {\bf 26}(1987)311-333, 335-355 

\bibitem{Rogula 99}D. Rogula,
Dynamics of magnetic flux lines and critical fields in high T\(_c\) superconductors, {\em J. Tech. Phys.} {\bf 40}(1999)383-406 

\bibitem{Rogula+Sztyren 2000}D. Rogula and M. Sztyren,
Equilibrium configurations of magnetic flux lines in strongly anisotropic superconductors, {\em J. of Tech. Phys.} {\bf 41}(2000)89-100 

\bibitem{Rogula+Sztyren 2001}D. Rogula and M. Sztyren,
Dynamics of flux lines in strongly anisotropic superconductors, {\em J. Tech. Phys.} {\bf 42}(2001)33-42 



\bibitem{Abrikosov 57}A.\,A. Abrikosov,
On magnetic properties of type II superconductors, {\em J. Exp. Th. Phys. (Russian)} {\bf 32}(1957)1442-1450 

\bibitem{Abrikosov book}A.\,A. Abrikosov,
{\em Fundamentals of the Theory of Metals,} North--Holland 1988  

\bibitem{Anthony+Seeger 73}K.\,H. Anthony and A. Seeger,
Eine nichtlineare Kontinuumstheorie des Flussliniengitters in Supraleitern, {\em Phil. Mag.} {\bf 28}(1973)1125-1148 

\bibitem{Bean 62}C.\,P. Bean,
Magnetization of hard superconductors, {\em Phys. Rev. Lett.} {\bf 8}(1962)250-253; {\em Rev. Mod. Phys.} {\bf 36}(1964)31-39 

\bibitem{Chapman+ 92}S.\,J. Chapman and S.\,D. Howson and J.R. Ockendon,
Macroscopic models of superconductivity, {\em SIAM Rev.} {\bf 34}(1992)529-560 

\bibitem{Chapman 2000}S.\,J. Chapman,
A hierarchy of models for type II superconductors, {\em SIAM Rev.} {\bf 42}(2000)555-598 

\bibitem{Kim+Stephen 69}Y.\,B. Kim and M.J. Stephen,
Flux flow and irreversible effects, {\em In: Superconductivity,} ed. H.Parks, New York 1969  

\bibitem{Laue book}M. von Laue,
{\em Theorie der Supraleitung,} Springer Verlag, Berlin-G\"{o}ttingen-Heidelberg 1949  

\bibitem{London+London 35a}F. London and H. London,
The electromagnetic equations of the supraconductor, {\em Proc. Roy. Soc. (London)} {\bf A149}(1935)71-88 

\bibitem{London I}F. London,
{\em Superfluids. I. Macroscopic theory of superconductivity,} Dover Publications, New York 1961  

\bibitem{Zhou book}
Shu-Ang Zhou, {\em Electrodynamic theory of superconductors}, Peter Peregrinus Ltd., London 1991  

\bibitem{Kosevitch 79}A.\,M. Kosevitch,
Crystal dislocations and the theory of elasticity, {\em Dislocations in Solids,} Vol. 1, ed. F.R.Nabarro, North--Holland 1979  

\bibitem{Kroner kontinuumstheorie}E. Kr\"{o}ner,
Kontinuumstheorie der Versetzungen und Eigenspannungen, {\em Ergebnise der Angewandte Mathematik} {\bf 5}, eds. L.Collatz, F.L\"{o}sch, Springer Verlag, Heidelberg 1958  

\bibitem{Kroner 81}E. Kr\"{o}ner,
Continuum theory of defects, {\em In: Physics of Defects,} Les Houches 1980, edited by R.Balian et al., North--Holland 1981  

\bibitem{Kroner 89}E. Kr\"{o}ner,
The differential geometry of elementary point and line defects in Bravais crystals, lecture at the Summer School on Topology, Geometry and Gauging, Jab{\l}onna 1989; {\em Int. J. Theor. Phys.} {\bf 29}(1990)1219-1237 


\bibitem{Mistura}L. Mistura,
Cartan connections and defects in Bravais lattices, lecture at the Summer School on Topology, Geometry and Gauging, Jab{\l}onna 1989; {\em Int. J. Theor. Phys.} {\bf 29}(1990)1207-1218 

\bibitem{Trzesowski 87}A. Trz\c{e}sowski,
Geometry of crystal structure with defects, {\em Int. J. Theor. Phys.} {\bf 26}(1987)311-333, 335-355 

\bibitem{Rogula 99}D. Rogula,
Dynamics of magnetic flux lines and critical fields in high T\(_c\) superconductors, {\em J. Tech. Phys.} {\bf 40}(1999)383-406 

\bibitem{Rogula+Sztyren 2000}D. Rogula and M. Sztyren,
Equilibrium configurations of magnetic flux lines in strongly anisotropic superconductors, {\em J. of Tech. Phys.} {\bf 41}(2000)89-100 

\bibitem{Rogula+Sztyren 2001}D. Rogula and M. Sztyren,
Dynamics of flux lines in strongly anisotropic superconductors, {\em J. Tech. Phys.} {\bf 42}(2001)33-42 


\end{document}